# Virtual Reality Alters Perceived Functional Body Size

Xiaoye Michael Wang [a]

Ali Mazalek [b]

Catherine M. Sabiston [a]

Timothy N. Welsh [a]

RUNNING HEAD:

[a.] Faculty of Kinesiology and Physical Education, University of Toronto, Toronto, ON, Canada

[b.] Synaesthetic Media Lab, Toronto Metropolitan University, Toronto, ON, Canada

Corresponding author: Xiaoye Michael Wang. Email: michaelwxy.wang@utoronto.ca

**Abstract**

Virtual reality (VR) introduces sensorimotor perturbations that may impact perception and action. The current study was designed to investigate how immersive VR presented through a head-mounted display (HMD) affects perceived functional body size using a passable aperture paradigm. Participants (n=60) performed an action task (sidle through apertures) and a perception task (adjust aperture width until passable without contact) in both physical, unmediated reality (UR) and VR. Results revealed significantly higher action and perceptual thresholds in VR compared to UR. Affordance ratios (perceptual threshold over action threshold) were also higher in VR, indicating that the increase in perceptual thresholds in VR was driven partly by sensorimotor uncertainty, as reflected in the increase in the action thresholds, and partly by perceptual distortions imposed by VR. This perceptual overestimation in VR also persisted as an aftereffect in UR following VR exposure. Geometrical modelling attributed the disproportionate increase in the perceptual threshold in VR primarily to depth compression. This compression, stemming from the vergence-accommodation conflict (VAC), caused the virtual aperture to be perceived as narrower than depicted, thus requiring a wider adjusted aperture. Critically, after mathematically correcting for the VAC's impact on perceived aperture width, the affordance ratios in VR became equivalent to those in UR. These outcomes demonstrate a recovered invariant geometrical scaling, suggesting that perception remained functionally attuned to action capabilities once VAC-induced distortions were accounted for. These findings highlight that VR-induced depth compression systematically alters perceived body-environment relationships, and thus functional body size. If uncorrected, this systematic perturbation could add device-specific learning demands that may undermine transfer of VR-based motor skill training to UR.

## 1. Introduction

If digital peripheral devices such as a mouse and a keyboard are the interface between users and a computing device, then the body can be considered as the interface between humans and the environment. In the long history of inquiries that endeavour to uncover the secret of human consciousness, the pivotal role of the body has been generally overlooked in comparison to that of the brain. However, one does not come to know the world by merely interpreting the sensory information with one's brain, but by *dwelling* in the world with one's body through intentional movement, skilled action, and embodied perception (Heidegger, 1962; Merleau-Ponty, 1945). In this context, the body is not an object, but the medium of disclosure through which the world becomes meaningful (Johnson, 2007). This embodied mode of knowing has profound implications for immersive virtual reality (VR), where the presence or absence of a body representation fundamentally shapes the user's experience of the virtual world (Anjos & Pereira, 2024; Heidicker et al., 2017; Murray & Sixsmith, 1999; Slater, 2018; Tang et al., 2004). By leveraging devices such as a head-mounted display (HMD), VR uses advanced motion tracking and display technologies to simulate the perceptual experience of the physical, unmediated reality (UR) (Wang & Troje, 2023, 2024). Because the body serves as an intrinsic scale by which organisms perceive and act on environmental properties specific to a certain task (J. J. Gibson, 1979), providing a coherent, full body representation that maps onto the user's movements is critical in VR for preserving body-environment calibration and enabling the prospective control of action.

Traditionally, capturing a user's full body movement and using it to animate a virtual avatar required the integration of additional hardware, such as marker-based (e.g., OptiTrack, Vicon) or markerless (e.g., Microsoft Kinect) motion capture systems (e.g., Sobota et al., 2016; Spanlang et al., 2010). Recently, with the rapid advancement of machine learning and computer vision, on-device, inside-out body tracking has grown in popularity and, more importantly, improved effectiveness and accuracy in enabling users to control a full-body avatar. This improved interaction fidelity, together with VR's flexibility in generating controlled, replicable training scenarios, has contributed to the growing use of VR for skill training (Xie et al., 2021). However, successful training in VR and its subsequent transfer to UR depend on whether the task-specific body-environment calibration learned in VR remains valid outside the headset. Recent studies have shown that HMD-based VR introduces visual and sensorimotor perturbations that degrade perceptuomotor performance in VR (Wang et al., 2024b, 2025) and have negative transfer to UR, that is, more errorful performance in UR after than before performance in VR (Wang et al., 2024a). If these perturbations systematically distort perceived body-environment relationships, users may learn device-specific compensations that do not generalize, or even have negative transfer, to UR. Accordingly, the current study examines how immersive VR affects perceived body-environment scaling (operationalized as perceived passability/functional body size) and whether these effects persist after returning to UR.

### 1.1 Affordances and Functional Body Size

The research question addressed herein was: *How do users perceive their body size in VR?* To begin answering this question, it is important to recognize that body perception is inherently relational: perceiving one's body does not occur in isolation, but rather within an environment populated by objects and people, along with various ways of interacting with them. From the perspective of ecological psychology, these interactions are described in terms of affordances, or the action possibilities that the environment offers to an organism given the organism's relevant action capabilities (J. J. Gibson, 1979; see also Turvey et al., 1981; Kohm et al., 2025). Affordances capture the functional relationship between the properties of the environment and those of the organism and can be directly picked up by the organism through the optical patterns available at a given time and place. Warren (1984) operationalized affordances as a $\pi$ ratio between a relevant environmental dimension ($E$) and the corresponding organism dimension ($O$) for a certain action:

$$\pi = \frac{E}{O} \quad \text{Equation 1}$$

This ratio defines the action boundary, or the point at which the environment no longer supports a particular action for a given individual, thereby requiring the individual to change the mode of the action to complete the task.

In a foundational study, Warren and Whang (1987) asked participants to walk through doorway-like openings, or apertures, that varied in width (the distance between the left and right edges of a door frame). The mode of action changed depending on the width of the aperture: when the opening was wide enough, participants would walk straight through, but when it was too narrow relative to their shoulder width, participants turned their shoulders to sidle through sideways instead. The width of the aperture at which participants turned their shoulder, or the critical aperture width, defines the action boundary between walking and sidling. By systematically manipulating aperture widths, the authors found that although the absolute widths at which participants altered their movement varied with body size (e.g., taller and larger individuals typically required wider apertures to walk straight through), the geometrical scaling between the critical aperture width ($A$) and shoulder width ($S$) remained invariant between small and large participants:

$$\pi_{critical} = \frac{A_{small}}{S_{small}} = \frac{A_{large}}{S_{large}} \quad \text{Equation 2}$$

Their findings showed that $\pi_{critical}$ was approximately 1.3, indicating that participants reliably began turning their shoulders when the aperture was about 1.3 times their shoulder width or less, regardless of individual shoulder width. Furthermore, participants were also asked to view the apertures from a stationary position at a 5 m distance and indicate whether they thought they could pass through without turning their shoulders. These perceptual judgments showed similar invariance, where the ratio between the judged threshold for passability and each participant's shoulder width remained consistent across individuals.

This invariant organism-environment scaling has also been demonstrated in other tasks, such as stair climbing (Warren, 1984), throwing (Zhu & Bingham, 2011), and reaching-to-grasp movements (Bingham, Snapp-Childs, et al., 2014; Wang & Bingham, 2019). Across these tasks, observers can accurately perceive the relationship between their own body dimensions and action capabilities in relation to the relevant environmental demands. On the one hand, the affordance ratio from the action task defines the action boundary and provides information about the individuals' body dimensions and action capabilities in relation to properties of the environment and the task demands, respectively. In fact, this ratio can serve as a functional measure of body size (Bingham, Pan, et al., 2014; Wang & Bingham, 2019). On the other hand, the affordance ratio from perceptual judgments reveals how individuals perceive dimensions of their body and action capabilities relative to the environment. The invariance of these ratios across individuals of different body sizes suggests that affordance perception is body-scaled and action-specific, rather than based on absolute dimensions. However, a critical question remains: how is the body-scaled affordance information specified through visual information?

In the context of passable apertures, Warren and Whang (1987) hypothesized that the participant's eye height was a factor that helped to relate the observer's shoulder width to the perceived aperture width (J. J. Gibson, 1979; Sedgwick, 1973; Sedgwick et al., 1980; Warren & Whang, 1987). To test this hypothesis, the authors presented apertures on a raised platform viewed through a reduction screen, which effectively lowered participants' perceived eye height without their awareness. This change led to lower perceptual judgment thresholds and a reduced affordance ratio (aperture width relative to shoulder width), suggesting that participants perceived their bodies to be smaller relative to the environment. In other words, altering the perceived spatial layout of the environment changed how participants perceived their own body dimensions in relation to it.

Figure 1 shows how optical information, measured in an angular reference frame, specifies the metric relationship between eye height and aperture width. The observer's eye height is denoted as $H$, which corresponds to a declination angle $\gamma$ on a horizontal ground surface. This angle $\gamma$ is the optical variable specifying eye height and distance (Messing & Durgin, 2005; Ooi et al., 2001; Renner et al., 2013). The aperture width is $W$ that subtends a visual angle $\alpha$. Using trigonometry, the following relationship can be established (Warren & Whang, 1987):

$$\frac{W}{H} = \frac{2\tan\frac{\alpha}{2}}{\tan\gamma} \qquad \text{Equation 3}$$

When the aperture was presented on a raised floor, the declination angle was reduced without altering the actual eye height $H$. Because this manipulation did not affect the visual angle $\alpha$ subtended by the aperture, Equation 3 can only remain valid if the perceived aperture width $W$ increases, which yields:

$$\frac{\widehat{W}}{H} = \frac{2\tan\frac{\alpha}{2}}{\tan\hat{\gamma}} \qquad \text{Equation 4}$$

Where $\hat{\gamma}$ is the reduced declination angle due to the raised floor whereas $\widehat{W}$ is the perceived aperture width based on the perturbed declination angle. This equality indicates that a smaller aperture width would be sufficient to produce the same perceived aperture width as in the baseline condition without the raised floor. This explains why perceptual judgments of the critical aperture width shifted accordingly.

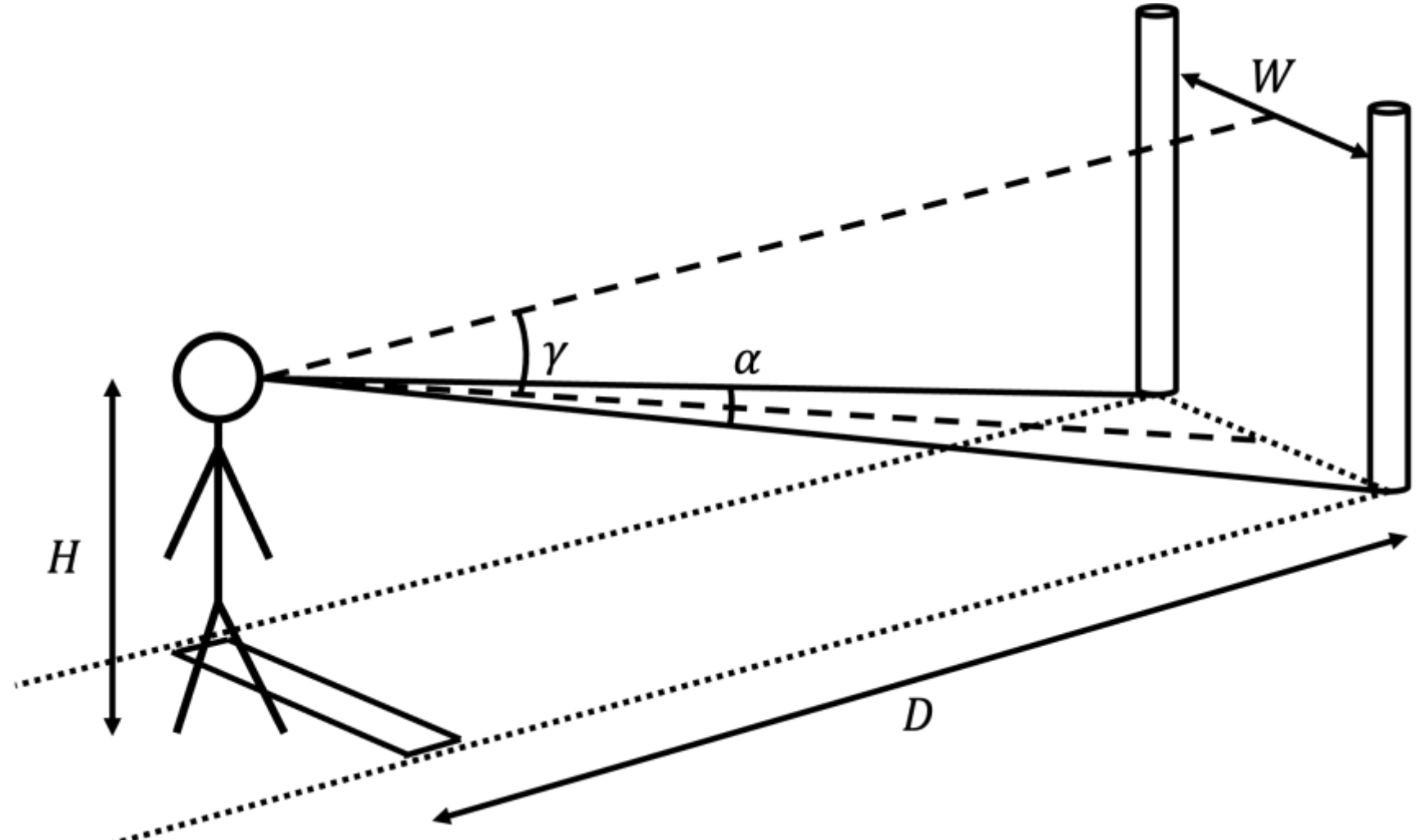

Figure 1. The scaling relationship between the observer's eye height and the width of a passable aperture. At eye height $H$, the observer stands distance $D$ away from the depicted aperture, corresponding to a declination angle $\gamma$. The minimum width of the aperture for the observer to pass without touching the aperture is $W$ that subtends a visual angle $\alpha$.

In sum, affordance analysis grounded in invariant body scaling provides a powerful framework for evaluating body perception in both physical and virtual environments. If perception remains attuned to action capabilities, then affordance ratios should remain invariant across contexts. Changing the individuals' action capabilities and perceived spatial extent can shift how they perceive the dimensions of their body in the environment, creating deviations from this invariance. Therefore, affordance analysis can be used as a diagnostic tool to identify the sources and consequences of perceptual distortions or sensorimotor uncertainty introduced by VR. In other words, testing whether invariant affordance scaling holds in VR offers critical insight into how virtual environments reshape the perceived relationship between body and world, and by extension, how they may alter users' perception of their own body size or capabilities.

## 1.2 Sensorimotor Perturbations in VR

The digital simulation of a 3D environment presented via a VR HMD introduces sensorimotor perturbations that can impact how users perceive and interact through their avatar. For instance, common HMDs are relatively heavy (Ito et al., 2021), have a restricted field of view (Knapp & Loomis, 2004; Willemsen et al., 2009), introduce motion-to-photon latency (Warburton et al., 2023), and lack appropriate proprioceptive feedback (Sra et al., 2019). Moreover, VR systems use digital displays to render different depth information to provide users a sense of being immersed in an environment different from the one they are physically located in, producing a sense of presence (Slater, 2009, 2018). For human observers, depth perception is governed by several mechanisms, including linear perspective (Saunders & Backus, 2006; Todorović, 2005; Wu et al., 2007), occlusion (Fischer et al., 2023; He et al., 2004; Shimojo et al., 1988), binocular disparity (Backus et al., 1999; Wang & Troje, 2023, 2024), and motion parallax (E. J. Gibson et al., 1959; Ono et al., 1988; Rogers & Graham, 1979). The two key types of depth information that distinguish the experience of using an immersive VR from simply interacting with 3D virtual environments on a flat screen are motion parallax and binocular disparity (Wang & Troje, 2023, 2024).

Of particular focus for the present paper is binocular disparity, a source of depth information generated by the slight offset between the retinal images of the two eyes resulting from their lateral separation (Howard et al., 1995; Julesz, 1971). To derive binocular disparity, the visual system has to generate and integrate signals from each eye to form a cohesive percept, as opposed to seeing double images (i.e., diplopia). In this process, the eyes rotate to converge (or diverge) on an object to align the retinal images so that the images fall on corresponding locations on the foveae to enable binocular integration in cortical visual centres. This process is called *vergence*. Simultaneously, when viewing objects at different distances, the ciliary muscles adjust the shapes of the lenses of the eyes to keep the retinal images in sharp focus. This process is termed *accommodation*. In UR, vergence and accommodation are tightly coupled, such that changes in one influence the other (Eadie et al., 2000; Hung, 1992; Hung et al., 1996). In most HMD-based VR systems, however, the lenses of the headset present images on the screen at a fixed focal distance. As a result, the eyes remain focused at a single, fixed distance, preventing natural adjustments in accommodation. In contrast, the eyes continue to adjust in orientation (vergence continuously occurs) as the user fixates on different virtual objects at varying depths and locations. The decoupling of static accommodation and dynamic vergence in HMD-based VR creates the *vergence-accommodation conflict* (VAC). There have been attempts to develop optical systems for VR HMDs that dynamically adjust the display's focal distance based on the user's fixation (e.g., Chang et al., 2018; Dunn et al., 2017; Kaneko et al., 2021; Liu et al., 2009). However, significant challenges remain in creating a consumer-level product that is both effective at supporting multiple focal distances and compact in form.

Based on a geometrical model from Wang and colleagues (Wang et al., 2024b, 2026) grounded in the neural coupling between vergence and accommodation, the constant accommodative demand can disrupt the accommodative vergence process, biasing the vergence angle inward closer to the display plane. This shift in vergence is predicted to cause a systematic misinterpretation of disparity information: even though the stereoscopic display presents binocular disparities consistent with the rendered 3D scene, the altered vergence angle leads the visual system to interpret these disparities as indicating shallower depth, resulting in depth compression (see also Singh et al., 2018; Swan et al., 2015). In support of the predicted depth compression, Wang et al. (2024a) found that the manual aiming movements of people in VR were shorter than those same movements in UR. Interestingly, this undershooting persisted when the user returned to move in the physical world, suggesting temporary shifts in the accommodative vergence response that persist beyond the HMD. The model of the VAC can account for these pointing errors in VR/AR (Neveu et al., 2016; Wang et al., 2024a; Yego et al., 2025).

In sum, although HMD-based VR can elicit a strong sense of presence, various factors, from HMD's mechanical constraints to the ways through which depth information is conveyed, can result in perceptual distortions and perturbed perception-action coupling, leading to reduced movement accuracy, greater movement variability, and slower movement execution (Wang et al., 2025). Because affordances are specified via optical patterns and are grounded in the relationship between the individual's action capabilities and environmental properties, such disruptions may lead users to recalibrate how they perceive their body's ability to act. In other words, the sensorimotor perturbations imposed by HMD-based VR can alter the perceived *functional body size*, that is, how large, capable, or effective the virtual body feels relative to task demands. From an affordance perspective, these differences would manifest as deviations from invariant affordance ratios, providing a diagnostic window into how VR systematically reshapes the perceived relationship between body and environment.

### 1.3 Does VR Affect Perceived Functional Body Size?

The present study used a passable aperture task to investigate whether sensorimotor perturbations in VR alter passability judgments, thereby revealing shifts in perceived functional body size. The present paper reports a subset of the data from a larger study (see Wang et al., in preparation) in which participants were asked to sidle through a series of apertures of different widths in UR and VR. The aperture in UR was created by placing two vertical wooden poles at different distances from each other. The wooden poles were placed in stands that held them upright, but the stands were not stable, so the poles could be easily knocked over if contacted. The VR environment was a digital twin of the physical room with upright wooden poles that could also be knocked over if contacted by the avatar. The main measure in the study was the width threshold, or the maximum width of the aperture between the poles, that the participants could sidle through without touching either end of the aperture.

Participants completed both action and perception tasks. In the action task, participants walked up to the physical or virtual poles and sidled through without knocking them over. The action task not only enables participants to calibrate optical information to guide action through perceptual feedback (poles falling over), which in turn supports more accurate perceptual judgments (Fajen, 2005, 2007), but also provides a functional measure of the participants' physical (and virtual) dimensions in relation to the task demand (Bingham, Pan, et al., 2014; Wang & Bingham, 2019). Due to factors such as restricted field of view (Gagnon et al., 2021) and a lack of proprioceptive feedback (Mestre et al., 2016), movement execution may be prone to increased sensorimotor uncertainty. This uncertainty can lead to greater movement variability, prompting participants to require wider apertures to successfully sidle through without contact. Accordingly, it was hypothesized that action thresholds would be higher in VR than in UR. In the perception task, an experimenter adjusted the aperture width until the participants thought they could sidle through without touching the aperture. It was hypothesized that, similar to the action threshold, the perceptual threshold would increase in VR compared to UR.

In addition to predictions about passability in VR, this study was also designed to determine if any perceptual distortions in VR impacted perception and action when participants returned to UR. On a neural level, VAC could temporarily alter the accommodative vergence response, leading to a persistent perturbation to binocular vision after prolonged VR use that could result in an adaptation aftereffect manifested as depth compression in UR (Wang et al., 2024a). This perceptual distortion could lead to participants perceiving their physical body as larger relative to the environment upon returning to UR. This has clear implications for VR-based training in tasks where clearance is critical (e.g., industrial maintenance in constrained spaces or emergency egress training), because users may learn device-specific safety margins that do not generalize to UR. To test this prediction, participants completed the perception task in UR both before and after performing the tasks in VR. If wider perceived apertures in VR shift perceived body size, then participants should judge that they require larger apertures in the physical world to pass through after VR exposure compared to before exposure.

## 2. Methods

### 2.1 Participants

Sixty (60) adults (33 females and 28 males, mean age = 22.46 years, SD = 4.05) participated in this study. All participants were neurologically healthy and had normal or corrected-to-normal vision. Experimental procedures were approved by the University of Toronto Research Ethics Board in accordance with the Declaration of Helsinki. All participants provided full and informed written consent before participation and received monetary compensation upon completion of the study.

### 2.2 Stimuli and apparatus

The experiment was executed using a desktop computer with an Intel Core i7-13700K Processor, 64 GB RAM, and an NVIDIA RTX 4080 Graphics Card. The experiment was conducted in a physical (UR) and a virtual (VR) environment. In UR, two wooden poles (190 cm in height, 4 cm in depth and width) supported by a wooden base, were placed 2.5 m away from the participant.

These wooden poles were manually moved by the experimenters to set the width of the aperture (Figure 2a). The virtual environment and its associated tasks were implemented using Unity and presented through an HTC VIVE Pro 2 VR HMD with a resolution of 2448 × 2448 pixels per eye, a combined 120° field of view, and a 90 Hz refresh rate. Participants responded with an HTC VIVE 2.0 Controller. The dimensions of the virtual environment were similar to those of UR (Figure 2b). Inside the virtual environment, two virtual versions of the wooden poles were placed 2.5 m away from the participant

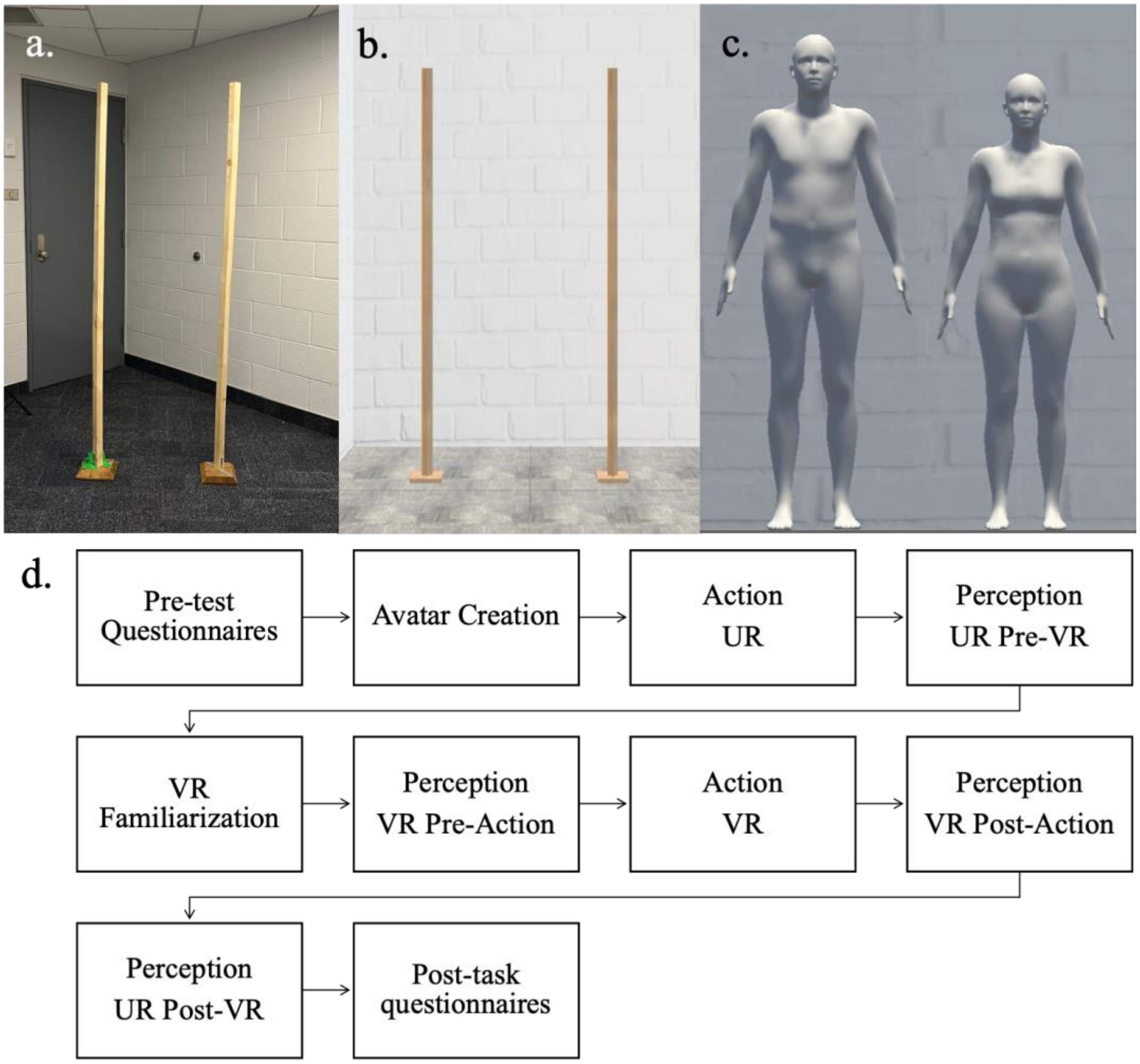

Figure 2. (a-b) Apertures used in the physical and virtual environments. (c) Custom avatars with a male and a female body in the virtual environment. (d) Experimental procedures. See text for details.

A custom virtual avatar was created for each participant (Figure 2c). Avatar creation was achieved using the Virtual Caliper (Pujades et al., 2019) for 3D body measurement and an avatar creation tool based on the SMPL (skinned multi-person linear) model (Loper et al., 2023). The Virtual Caliper uses two HTC VIVE controllers to take 3D measurements of the participant's height, arm span (finger to finger), and inseam height. These measurements, combined with the participant's weight, were used to tune the body shape parameters in the SMPL model, which maps pose $\theta$ and shape $\beta$ to produce a triangulated mesh, $\mathcal{M}$, that represents the avatar. This tool allows the creation of realistic avatars that reflect the user's body dimensions. The avatar's texture was a uniform grey for all participants. As a part of a bigger study, the avatar's body size was manipulated by modifying the input weight to be the same (n=20), 20% more (n=20), or 20% less (n=20) of participants' physical weight during the avatar creation process. This manipulation was aimed at addressing a different research question and thus will be reported in another paper (Wang et al., in preparation). While the Virtual Caliper offers as many as 6 measurements as input, including hip width and arm length, pilot testing showed that restricting the hip width while using a different weight for the avatar from the participant's original weight would yield unrealistic body shapes. Therefore, only 4 measurements (i.e., height, weight, arm span, fingers, and inseam height) were used as inputs to create the avatars. Participants controlled the avatar via five VIVE Trackers 2.0 – one tracker affixed to their hands, feet, and the waist using Velcro straps. Inverse kinematics protocols were used to animate the avatar using positions and orientations from the trackers and the HMD.

## 2.3 Procedures

Prior to the beginning of each session, experimenters calibrated the location of the floor using SteamVR to ensure consistent height between VR and UR. Figure 2d shows the experimental procedures for the entire study. Participants were instructed to wear form-fitting clothing before arriving at the lab. The experimenters first introduced the purpose and general procedures of the study to the participant, who subsequently provided informed consent before the experiment. Participants completed a series of pre-test questionnaires, including Body and Appearance Self-Conscious Emotions Scale (BASES; Castonguay et al., 2014), Body Appreciation Scale 2 (BAS-2; Tylka & Wood-Barcalow, 2015), Functionality Appreciation Scale (FAS; Alleva et al., 2017), Modified Weight Bias Internalization Scale (m-WBIS; Pearl & Puhl, 2014)), and Positive and Negative Affect Schedule (PANAS; Watson et al., 1988). Detailed descriptions of and participants' responses to these questionnaires will be reported in a separate paper. Then, the experimenter took the participants' body measurements, including weight, height, inseam height, arm span, and arm length, which were then used to create a custom avatar.

After these initial steps and avatar setup, participants performed a series of action and perception tasks with the apertures in UR and VR using a within-subject design. For the action tasks, participants started 2.5 m away and then walked towards the aperture, turned sideways, and sidled through it. The experimenter recorded whether the participants came into contact with the aperture, where one or both poles may fall over after contact in the physical or virtual environments. A total of 18 aperture widths were presented, ranging from 20 to 60 cm with a 5 cm increment. The apertures were presented sequentially in one of two orders: (1) ascending (20 cm to 60 cm) followed by descending (60 cm to 20 cm), or (2) descending (60 cm to 20 cm) followed by ascending (20 cm to 60 cm). The order was counterbalanced across participants. At the beginning of each trial, participants closed their eyes, and the experimenter manually adjusted the aperture width in UR, or the aperture width was set automatically via a program in VR.

Participants performed the perception task 2.5 m away from the aperture. Initially, the perception task was designed as a method of limits procedure, presenting discrete aperture widths in both ascending and descending series (order counterbalanced across participants) spanning 20 to 60 cm in 5-cm increments (9 aperture widths per series; 18 trials total). Pilot testing indicated that this discrete yes/no approach was inefficient, as it required responses to many apertures that were clearly passable or clearly impassable, while still limiting threshold resolution due to the step size. Accordingly, the method of adjustment was adopted for the perception task, which allows participants to efficiently converge on their passability boundary and yields higher-resolution perceptual threshold estimates with fewer trials. Specifically, participants adjusted the distance between the two poles to the narrowest distance that they judged they could sidle through without contacting either pole. In UR, participants instructed an experimenter to manually adjust the aperture width by moving one of the poles to either bring them closer (reducing the aperture width) or move them farther away from each other (increasing the aperture width). An experimenter was always present close to the aperture and participants were instructed to focus on the width of the opening between the two poles. Across data collection, this adjustment was performed by four different experimenters with different heights and body sizes. The location of the starting poles changed on each trial, and the direction of the adjustment was counterbalanced between trials. Participants could stop the movement of the pole at any time to indicate the appropriate aperture width and were allowed to fine-tune their response before confirmation. In VR, participants directly adjusted the aperture width using a VIVE controller, pressing the up/right button increased the aperture and pressing the down/left button decreased the aperture. This task was repeated ten times for each perception task, and the adjusted widths were recorded.

Participants first performed the action task, followed by the perception task in UR. These "UR Pre-VR" trials were recorded to obtain a baseline measure of the participants' action and body size perception in UR prior to exposure to VR and the avatar. The action task was administered first to enable participants to calibrate dimensions of their physical body to the task in UR. Critically, the resulting action threshold was used as a functional measure of participants' body size (Bingham, Snapp-Childs, et al., 2014). The perception task provided a baseline measure of calibrated perceived body size for sidling through an aperture in UR. The initial action-perception sequence in UR deviates from the classical affordance paradigm, which also includes a perception task before the action task. Previously, the perception-action-perception sequence is intended to test whether performing the action could help observers to calibrate their action capabilities (e.g., body size) to the task demands (e.g., Fajen, 2008; Rieser et al., 1995; Withagen & Michaels, 2002). The primary purpose of the current study, however, focused on the contrast in performance between UR and VR, instead of the affordance for sidling through an aperture itself, where the latter has been well documented (e.g., Franchak, 2020; Franchak & Adolph, 2014). Therefore, to simplify experimental design and reduce session time, the baseline perception task in UR was not included.

Once the UR Pre-VR action and perception tasks were completed, participants were fitted with the HMD and trackers. Inside the virtual environment, participants stood in front of a virtual mirror and performed a series of 22 movements to familiarize themselves with their new virtual body and obtain a sense of embodiment and agency over the avatar. These familiarization movements included touching their head, lifting their arms, and jumping up and down. Note that the virtual apertures were not present during this initial familiarization procedure. After completing these movements, participants performed the perception ("VR Pre-Action"), action ("VR-Action"), and perception ("VR Post-Action") tasks in a sequential order in the virtual environment. The initial VR Pre-Action perception task was conducted to determine how participants perceived their avatar's body size without calibration from the action task in VR. The action task was conducted next to obtain participants' functional body size in VR and enable participants to calibrate their avatar's body size based on the unique task environment in VR. The virtual poles would fall over and make a noise if the avatar came into contact with them. The VR Post Action perception task was completed to determine if performing the action task in VR helped the participants calibrate their perceived virtual body size.

After completing the perception-action-perception task sequence in VR, participants removed the HMD and trackers and performed the perception task again in UR. The "UR Post-VR" perception task was conducted last to determine if the prolonged use of VR HMD impacts perceived functional body size in UR compared to the pre-VR measure. The inclusion of this task was intended to test the *a priori* hypothesis based on the adaptation aftereffect of VAC, where vergence is temporarily drawn inward upon returning to UR, resulting in depth compression that may affect perceptual threshold.

Finally, participants completed a set of post-test questionnaires, including PANAS (Watson et al., 1988), the Avatar Embodiment Questionnaire (AEQ; Peck & Gonzalez-Franco, 2021), Slater-Usoh-Steed Scale (SUS; Slater et al., 1998; Usoh et al., 1999), and Simulator Sickness Questionnaire (SSQ; Kennedy et al., 1993).

**2.4 Statistical Analysis**

The current paper focuses on the effect of perceptual distortions in HMD-based VR on perceptual body size and, therefore, only data from the affordance tasks were analyzed. These analyses were conducted without considering the avatar's body size manipulation. Statistical analysis was performed in R (R Core Team, 2020) using the *afex* (Singmann et al., 2015) and *emmeans* (Lenth, 2022) packages.

For the action task, psychometric curves were fitted to the binary passability data as a function of aperture width using the Wichmann and Hill model (Wichmann & Hill, 2001). The action threshold was defined as the mean (i.e., the point of subjective equality, or PSE) of the fitted curve. A one-factor repeated-measures analysis of variance (ANOVA) was used to evaluate the effect of Modality (2 levels: UR and VR) on action threshold. For the perception task, the perceptual threshold was calculated as the mean adjusted aperture width for each task. A one-factor repeated-measures ANOVA was used to evaluate the effect of Task (4 levels: UR Pre-VR, VR Pre-Action, VR Post-Action, UR Post-VR) on perceptual threshold. Greenhouse-Geisser correction was applied if analyses of sphericity revealed that the data violated the sphericity assumption, as indicated by degrees of freedom values reported with decimals. Finally, affordance ratios were computed as the division between perceptual threshold and action threshold for UR, VR Pre-Action, and VR Post-Action. Although the perception and action tasks differed procedurally (continuous adjustment *vs.* discrete yes/no responses), this feature does not complicate the affordance-ratio calculation because the action threshold was estimated as a continuous value by fitting a psychometric function to the binary passability responses. Thus, both perceptual and action thresholds entering the affordance ratio were continuous threshold estimates. These ratios were submitted to a repeated measures ANOVA with Task as a within-subject factor (3 levels: UR Pre-VR, VR Pre-Action, VR Post-Action).

Across all analyses, statistical significance was evaluated at $\alpha = 0.05$, and all hypothesis tests were two-tailed. Significant effects were further examined through post hoc pairwise comparisons using Tukey's HSD adjustment, yielding *p* values corrected for multiple comparisons. Effect sizes were reported as partial eta squared ($\eta_p^2$) for repeated-measures ANOVAs and Cohen's d for post hoc comparisons, with 95% confidence intervals where applicable.

## 3. Results and Discussion

### 3.1 Action and Perceptual Thresholds

Figure 3a shows the mean action and perceptual thresholds in UR and VR. For the action threshold, a repeated-measures ANOVA showed a significant effect of Modality, $F(1, 59) = 92.94, p < 0.001, \eta_p^2 = 0.61$. The action threshold in VR (mean = 35.29 cm, SE = 0.65) was significantly greater than that in UR (mean = 28.12 cm, SE = 0.47; mean difference = 7.16 cm, SE = 0.74). The significantly larger action threshold in VR than in UR supports the hypothesis that sensorimotor perturbations in VR lead participants to require wider apertures to sidle through without contact. Specifically, due to factors such as a restricted field of view and a lack of proprioceptive feedback, participants tended to misjudge their body positioning and left excessive space in front of or behind their bodies, resulting in collisions with one or both poles.

For the perceptual threshold, there was a significant effect of task, $F(2.15, 126.87) = 197.54, p < 0.001, \eta_p^2 = 0.77$. Post hoc pairwise comparisons showed significant differences between all task conditions (Figure 3a). Noticeably, consistent with the action thresholds, there was a significantly larger perceptual threshold when participants went from UR (UR Pre-VR: mean = 31.23 cm, SE = 0.66) to VR (VR Pre-Action: mean = 46.56, SE = 1.19; mean difference = 15.33 cm, SE = 0.83, $p < 0.001$). This increase occurred after participants completed the initial movement task to become familiar with embodying their avatar, but before they were able to calibrate the perceived virtual body size to the task demands via the action task in VR. Therefore, although participants had familiarized themselves with the virtual environment and embodied the avatar by this stage, the elevated perceptual thresholds likely reflect a lack of calibration to their actual action capabilities, which had not yet been grounded through task performance.

Following the action task in VR, the perceptual threshold decreased significantly from VR Pre-Action to VR Post-Action (VR Post-Action: mean = 43.60 cm, SE = 1.07; mean difference = 2.96 cm, SE = 0.85, $p < 0.001$). This decrease in perceptual threshold suggests that actively performing the task helped participants calibrate their perceptual judgments based on the functional size of their virtual body relative to the task demands. However, despite this significant decrease, the perceptual threshold in VR was still significantly greater than that in UR (UR Pre-VR vs. VR Post-Action: mean difference = 12.37 cm, SE = 0.78, $p < 0.001$). In other words, although action experience in VR facilitated recalibration of body-environment scaling, the continued overestimation of required aperture width suggests that additional perceptual or environmental factors may have influenced participants' judgments.

Finally, comparisons between the perceptual threshold in VR Post-Action and the threshold upon returning to the physical world showed a significant decrease from VR to UR (UR Post-VR: mean = 33.75 cm, SE = 0.73; mean difference = 9.85 cm, SE = 0.57, $p < 0.001$). However, this reduction did not revert the perceptual threshold to the pre-VR level because the threshold in UR Post-VR was still significantly greater than that in UR Pre-VR (mean difference = 2.52 cm, SE = 0.45, $p < 0.001$). Together, these findings suggest that experience in VR induced an aftereffect in participants' perceived functional body-environment scaling, such that their perceptual judgments in the physical world remained altered even after returning from the virtual environment.

In sum, these analyses revealed 4 main findings: 1) there was an increase in the action threshold from UR to VR, indicating reduced action precision in VR; 2) the perceived passible aperture threshold was larger in VR compared to UR, reflecting an overestimation of the space needed for successful passage; 3) perceptual judgments in VR became more aligned with action capabilities after participants performed the task demonstrating experience in VR supported perception-action calibration; and, 4) the increased perceptual aperture threshold in VR persisted upon returning to UR, signaling a shift in perceived body-environment scaling.

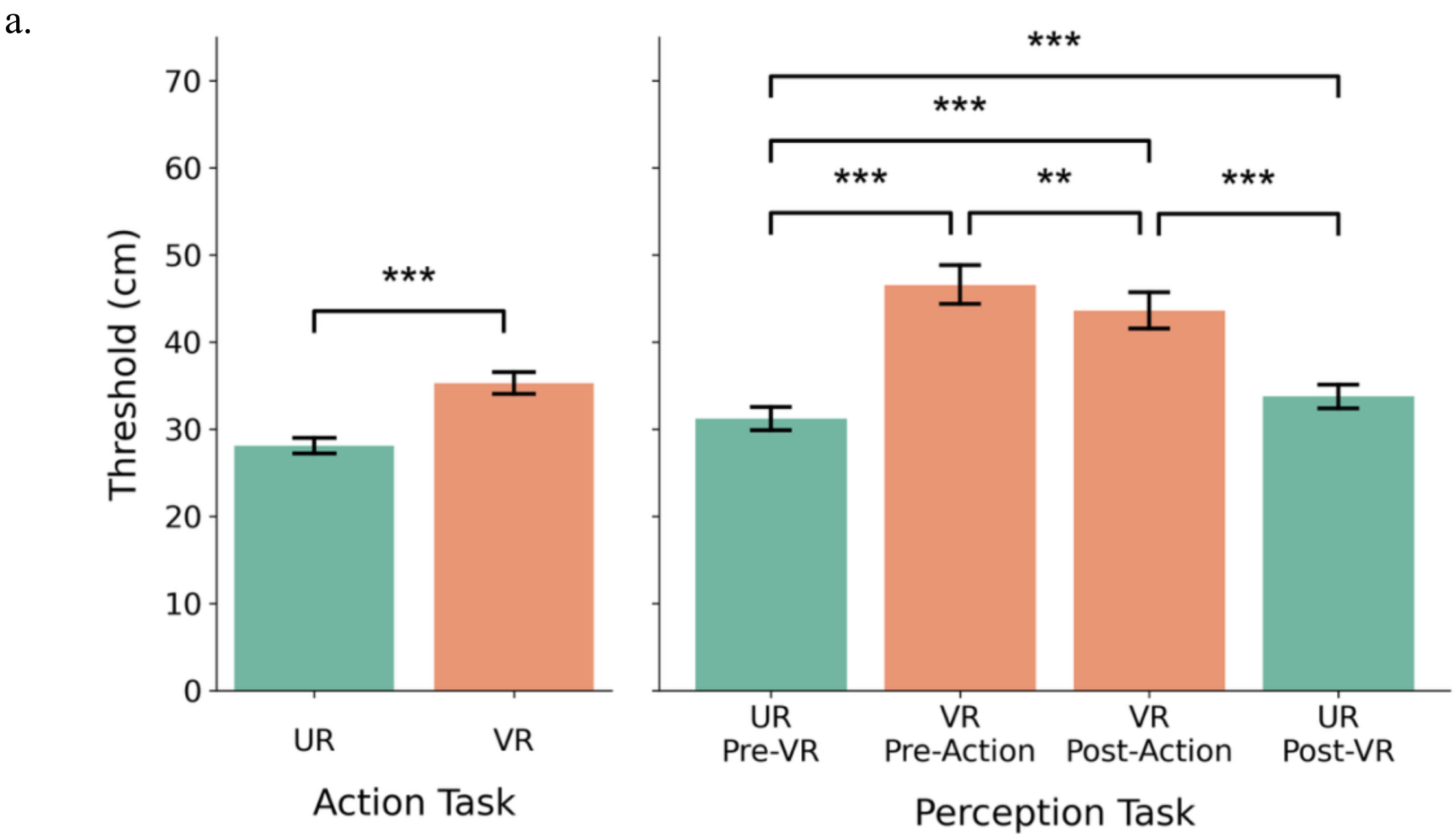

b.

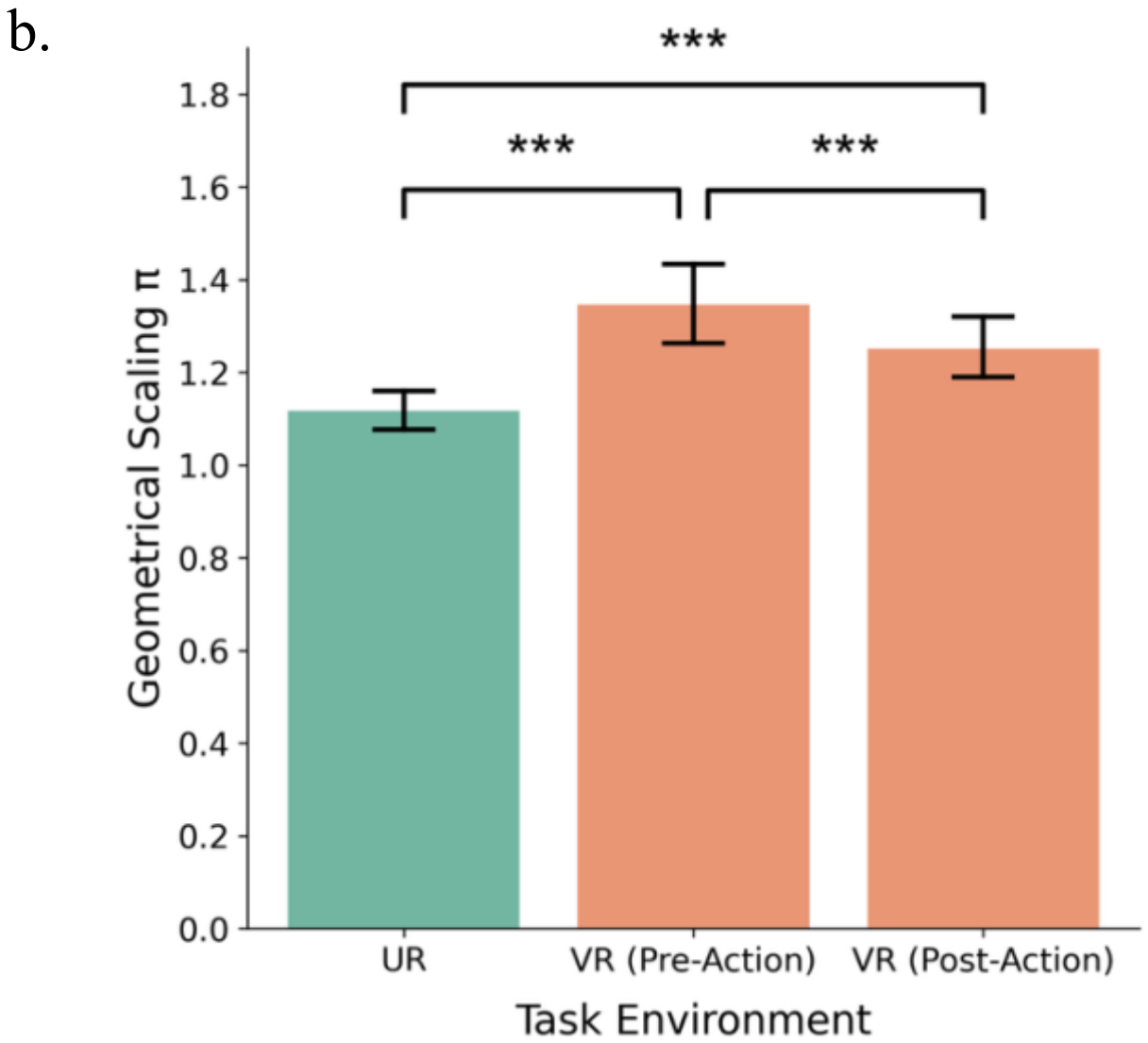

Figure 3. (a) The mean action and perceptual thresholds in UR and VR. (b) The mean geometrical scaling as the ratio between the perceptual and action thresholds in UR and VR (Pre- and Post-Action). Error bars represent the 95% confidence intervals. *: $p$ < 0.05; **: $p$ < 0.01; ***: $p$ < 0.001.

### 3.2. Affordance Ratios

The substantial increase in perceptual threshold when moving from UR (UR Pre-VR perception task) to VR (VR Pre-Action perception task) mirrored the increase in the action threshold (UR Action vs. VR Action). This increased perceptual threshold could reflect the uncertainty associated with VR that was similarly manifested in the action task. However, if this uncertainty were the sole source of such an increase, then the geometrical scaling between the participant's perceptual judgment and their action capability should remain invariant across modalities. Affordance ratios were computed and analyzed to evaluate this prediction.

The ANOVA revealed a significant effect of task on the affordance ratio, $F(1.64, 96.66) = 29.70, p < 0.001, \eta_p^2 = 0.34$. Post hoc pairwise comparisons showed statistically significant differences between every task pair. The affordance ratio in UR was smaller (mean = 1.12, SE = 0.023) compared to the ratio in VR before the action task (mean = 1.35, SE = 0.043; mean

difference = 0.23, SE = 0.036, $p < 0.001$). The relatively high affordance ratio in VR before the action task suggested larger safety margins in participants' perceptual judgment, reflecting the aforementioned sensorimotor uncertainty. Performing the task in VR resulted in a reduced affordance ratio (VR Post-Action: mean = 1.25, SE = 0.034; mean difference = 0.094, SE = 0.024, $p < 0.001$) compared to the affordance ratio prior to performing the action task in VR. This reduction indicated that participants could utilize the feedback from performing the action task to recalibrate their avatar's dimensions to the task demands, yielding more accurate perceptual judgment. Nevertheless, even after performing the action task, the affordance ratio in the VR Post-Action task was still significantly greater than the baseline affordance ratio in UR prior to VR experience (mean difference = 0.13, SE = 0.029, $p < 0.001$). This persistent offset in the affordance ratio implies that certain perceptual distortions, likely affecting visual processing but not action performance, continued to influence participants' judgments, resulting in elevated perceptual thresholds despite successful action calibration.

## 4. Perceptual Distortions and Passability Judgment

The increase in perceptual thresholds and affordance ratios as participants changed from UR to VR suggests that perceptual distortions in VR lead participants to perceive the virtual aperture as narrower, thereby requiring a wider aperture to sidle through. In Warren and Whang (1987), the authors manipulated the effective eye height to alter participants' perceptual thresholds for passing an aperture. Due to the perceptual distortions imposed by VR HMD, such as compression along the depth (Renner et al., 2013; Wang et al., 2024b) and even frontal dimensions (Kelly et al., 2015), perturbations to the perceived aperture extents along different dimensions (i.e., vertical, lateral, and in depth) could have contributed to the narrower perceived aperture width.

Based on Equation 3 and Equation 4, there is a positive scaling between the perceptual threshold and the declination angle:

$$\frac{\widehat{W}}{W} = \frac{\tan\hat{\gamma}}{\tan\gamma} \qquad \text{Equation 5}$$

If the increased perceptual threshold in VR was due to vertical distortions that altered effective eye height, then changes in perceptual thresholds can reveal the magnitude of the distortion. The ratio of the average perceptual aperture thresholds in VR (Pre-Action) and UR was 46.56 cm / 31.23 cm = 1.49. Therefore, the ratio between their corresponding declination angles is

$$\frac{\tan\hat{\gamma}}{\tan\gamma} = 1.49 \qquad \text{Equation 6}$$

Where declination angle $\gamma$ depends on the observer's eye height $H$ and the distance $D$ (Figure 1):

$$\tan\gamma = \frac{H}{D} \qquad \text{Equation 7}$$

Therefore, if changes in the declination angle were due to changes in eye height, the following can be established:

$$\widehat{H} = 1.49 \cdot H \qquad \text{Equation 8}$$

That is, the perturbed eye height in VR had to be 46% taller than the physical eye height to yield the observed increase in perceptual threshold in VR. To put into perspective, a person of height 1.7 m would stand 2.48 m in VR. Because the floor height of the VR HMD was always calibrated to the physical floor height before each session, such a noticeable height perturbation would be improbable.

Another potential source of perturbation is the lateral compression. If the perceived lateral extent is compressed in VR, participants would need to increase the physical width of the aperture to perceive it as sufficiently wide. Given the average perceptual thresholds in UR and VR, the lateral compression would need to be 31.23 cm / 46.56 cm = 0.67; that is, an object of length 30 cm would be perceived to be 20.1 cm in the virtual environment. While Kelly et al. (2015) reported lateral compression of a similar magnitude based on a blind walking task, the virtual environment that yielded this finding in their study was visually impoverished, with a pixelated grass texture against a gray, homogeneous backdrop. In the same study, a different, visually richer environment (i.e., a well-lit classroom) did not yield a similar lateral compression. Compared to the setup in Kelly et al. (2015), the environment used in the current study was more similar to the richer environment that did not yield lateral compression than to the impoverished environment that did. Therefore, it is also improbable that lateral compression was responsible for the greater perceptual threshold in VR.

After eliminating the potential influence of vertical and lateral perturbations, the remaining alternative explanation for the increased perceptual threshold in VR was depth compression. Numerous studies have shown depth compression in VR (Huang et al., 2021; Wang & Troje, 2023, 2024; for a review, see Renner et al., 2013) due to factors such as HMD weight (Buck et al., 2018), limited field of view (Willemsen et al., 2009), display resolution (Jää-Aro & Kjelldahl, 1997), and VAC (Batmaz et al., 2022, 2023; Singh et al., 2018; Swan et al., 2015; Wang et al., 2024b, 2026). To extrapolate the relationship between the perceptual threshold based on the aperture width and perceived depth, it is critical to differentiate between what is perceived in the perceptual space and what is depicted in the visual environment. According to Wang and Troje (2024), the spatial relationship between the visual environment $\Phi$ and the perceptual space $\Psi$ can be captured via a series of transformations $f$ specified by the available visual information:

$$\Psi = f(\Phi) \qquad \text{Equation 9}$$

In the visual environment, the aperture was 2.5 m from the observer, and the perceptual threshold was measured based on the distance between the two poles in the same visual environment. If the observer perceived the aperture to be at where it was depicted, the perceived aperture width specified by the visual angle $\alpha$ in the perceptual space would be equivalent to the measured aperture width $W$ in the visual environment (Figure 4). If the aperture was perceived to be at a closer, unknown distance, $d$, the visual angle $\theta$ corresponding to the optimal aperture width $W$ would be larger. Specifically, based on Equation 7, the perceived aperture distance $d$ is smaller than the depicted distance $D$, resulting in a larger perceived declination angle $\hat{\gamma}$:

$$\tan \hat{\gamma} = \frac{H}{d} \qquad \text{Equation 10}$$

According to Equation 4, a greater declination angle corresponds to a smaller perceived aperture width $\widehat{W}$. This relationship leads to the prediction that, if depth compression occurs and an aperture appears closer than it actually is, this aperture will also be perceived as narrower than it is intended to be presented. Consequently, the measured aperture width in the visual environment $W'$ would also be larger since the rendered aperture was at a farther location than the perceived aperture. Based on Equation 9, the recorded aperture width $W'$ in the visual environment has to be mapped onto the perceptual space to obtain the effective perceptual threshold:

$$\Psi(W, d) = f\big(\Phi(W', D)\big) \qquad \text{Equation 11}$$

Where $\Phi(W', D)$ corresponds to properties of the depicted virtual environment, $\Psi(W, d)$ is what the observer perceives, and $f$ is the geometrical transformations based on the visual information presented through a VR HMD.

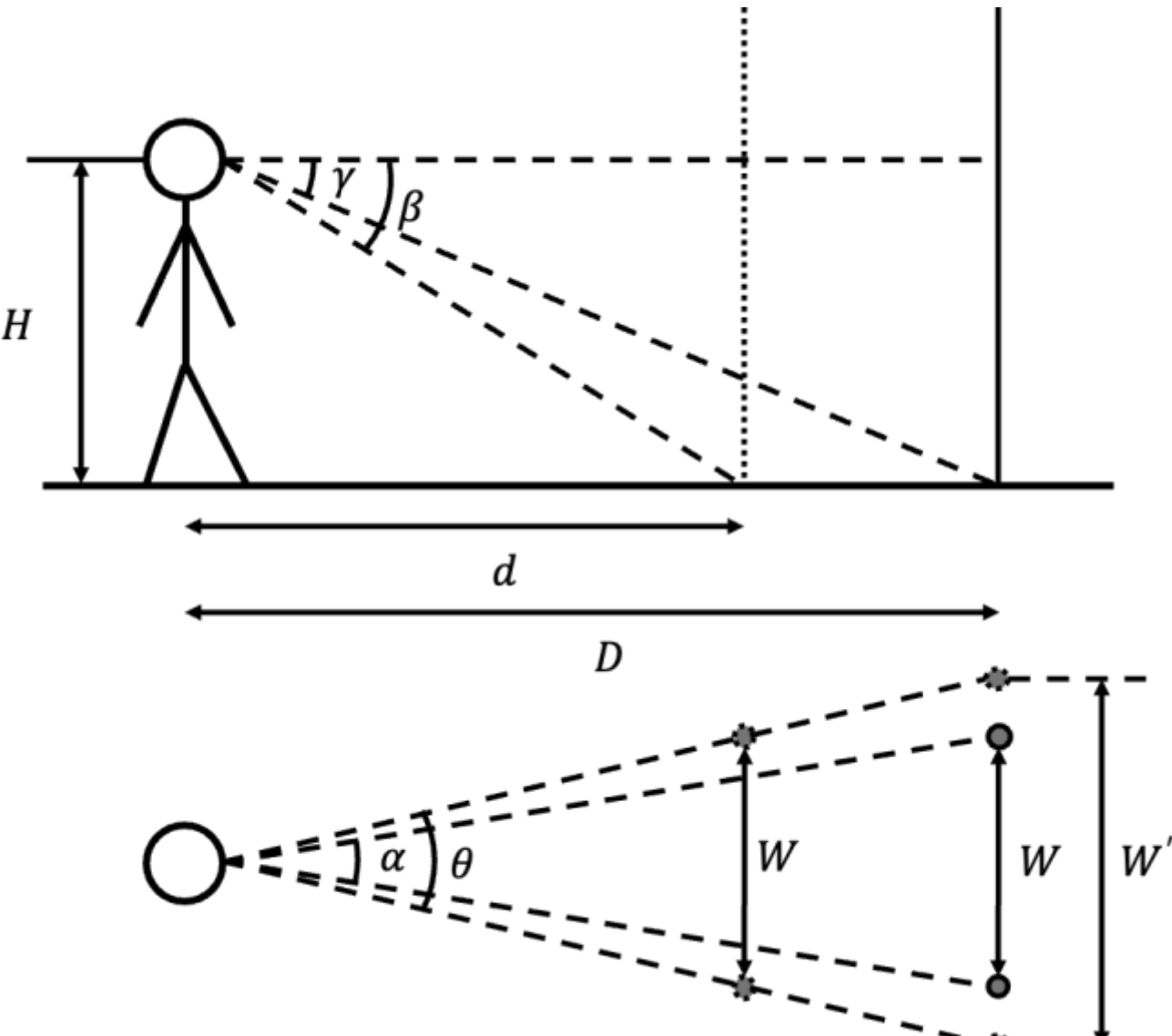

Figure 4. The scaling relationship between the observer's eye height and the width of a passable aperture from a profile (top) and an aerial (bottom) view. At eye height $H$, the observer stands distance $D$ away from the aperture (solid vertical line), corresponding to a declination angle $\gamma$. The minimum width of the aperture for the observer to pass without touching the aperture is $W$ that subtends a visual angle $\alpha$. With distance compression in VR, the aperture is perceived to be closer than it is depicted (dotted vertical line) at distance $d$ with a declination angle of $\beta$. Because the distance compression does not affect perceived dimensions on a frontoparallel plane, the observer still requires an aperture width of $W$ to pass. However, due to a shorter distance, the visual angle subtended by the aperture, $\theta$, is larger than the original visual angle $\alpha$. Maintaining the visual angle of the aperture, at the aperture's depicted distance $D$, the depicted aperture width is $W'$, which is larger than the width that the observer would require to pass the aperture.

Assuming the perceptual perturbations imposed by VR do not affect the perceived dimensions on a frontoparallel plane (Lind, 1996; Shapiro et al., 1995; Wang et al., 2018, 2020), the measured aperture width $W'$ in the visual environment should correspond to the aperture at distance $D$ that subtends the same visual angle $\theta$ (Figure 4). Using similar triangles, the following relationship can be established:

$$\frac{D}{W'} \propto \frac{d}{W} \Leftrightarrow W' \propto \frac{D}{d} W \qquad \text{Equation 12}$$

Note that this proportional form allows for the influence of other unmodeled factors (e.g., the sensorimotor uncertainty) that could impact perceptual judgment of aperture, rather than assuming a strict, deterministic relationship. Equation 12 suggests that the perceptual threshold $W'$ (i.e., what was measured in VR) is proportional to the perceived optimal aperture width $W$ (i.e., what the observer perceived) scaled by the ratio between the depicted and perceived distance, $D/d$. If the perceptual threshold in UR was equivalent to that in VR, the significant increase of the measured threshold from the physical to virtual environment indicated that $W' > W$, that is, $\frac{D}{d} > 1$ or $D > d$. In other words, the increased perceptual threshold from the physical to virtual environments could in part be attributed to distance compression.

### 4.1 The Vergence-Accommodation Conflict

After establishing that distance compression in VR could yield an increased perceptual threshold of passable aperture width, the subsequent question should focus on the mechanism of this distance compression. Wang and colleagues (2024b, 2026) presented a geometrical model to predict the impact of VAC on perceived distance. The authors argued that, given the bidirectional coupling between vergence (change in eye position) and accommodation (change, or lack thereof, in lens shape), the fixed accommodation distance imposed by the HMD pulls fixation inward, resulting in a positive offset to the vergence angle, which subsequently affects the binocular viewing geometry that specifies depth. Let the unperturbed binocular viewing geometry be:

$$\delta = \phi - \rho \qquad \text{Equation 13}$$

Where $\delta$ is the binocular disparity, $\phi$ is the fixation angle, and $\rho$ is the visual angle subtended by one of the poles of the aperture. Because of VAC, an offset $\beta$ is added to the vergence angle (Figure 5):

$$\hat{\phi} = \phi + \beta \qquad \text{Equation 14}$$

However, since the binocular disparity specified by the VR displays remains the same, the impact of the vergence offset is applied to the visual angle corresponding to the aperture:

$$\delta = (\phi + \beta) - \hat{\rho} \qquad \text{Equation 15}$$

Where $\hat{\rho}$ is the perturbed visual angle of the aperture in the perceptual space, which yields:

$$\hat{\rho} = (\phi - \delta) + \beta \qquad \text{Equation 16}$$

In other words, the perturbed visual angle subtended by the aperture increased by the same amount as how VAC pulls the vergence angle inward:

$$\hat{\rho} = \rho + \beta \qquad \text{Equation 17}$$

The visual direction of the aperture was specified by the display and should remain the same despite VAC. Therefore, the position of the perceived aperture is constrained to the line between the original aperture and the cyclopean eye (see Figure 4).

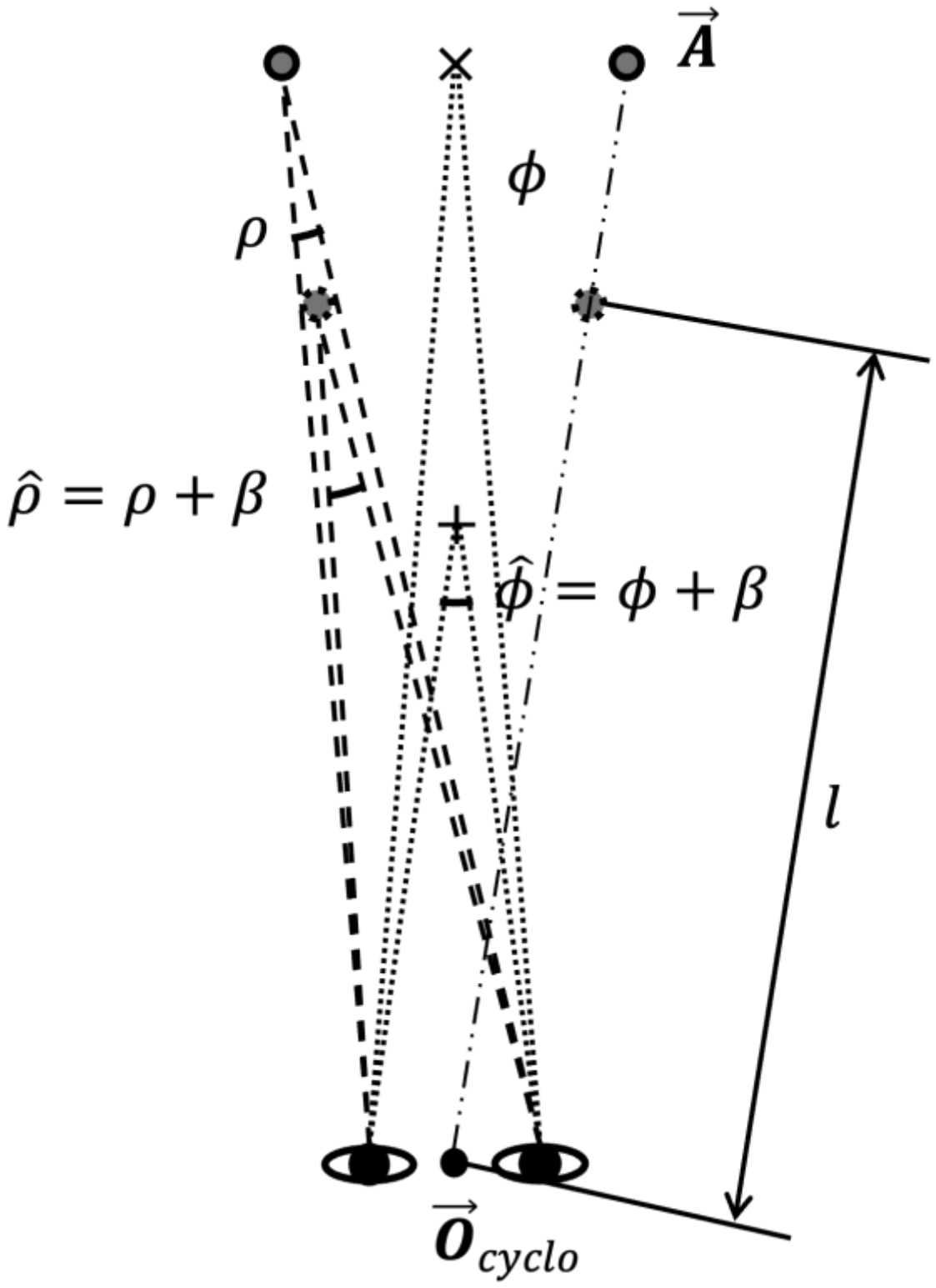

Figure 5. An illustration of the effect of the vergence-accommodation conflict (VAC) on perceived aperture width. Without the influence of VAC, the fixation angle is $\phi$ (× symbol), whereas the visual angle subtended by a pole of the aperture is $\rho$ (gray circles with solid outline). With VAC, the perturbed vergence angle $\hat{\phi}$ is drawn inward (+ symbol). Due to the fixed binocular disparity specified by the VR displays, the visual angle corresponding to the aperture $\hat{\rho}$ also increases by the same amount $\beta$. Because VAC does not affect the visual direction of the pole, the perceived

aperture remains constrained to the visual direction of the unperturbed aperture. At a closer perceived distance due to increased visual angle, the width of the aperture specified by binocular disparity is reduced.

As illustrated in Figure 5, the depth compression resulting from VAC also pulls the outer edges of the aperture inward, resulting in a narrower perceived aperture. To derive the perceived aperture locations, the perturbed visual angle of the aperture $\hat{\rho}$ and the interpupillary distance (IPD) are used to compute the Euclidean distance $l$ between the cyclopean eye and the perceived aperture location

$$l = \frac{IPD/2}{\tan\hat{\rho}} \qquad \text{Equation 18}$$

Then, the location of the perceived aperture can be found along the line between the cyclopean eye and the original target location $\vec{\boldsymbol{A}}$, with a distance $l$ from the cyclopean eye $\vec{\boldsymbol{O}}_{cyclo}$. Using parametric line equations, the perceived aperture location $\vec{\hat{\boldsymbol{A}}}$ can be computed as:

$$\vec{\hat{\boldsymbol{A}}} = \vec{\boldsymbol{O}}_{cyclo} + \left(\frac{IPD/2}{\tan(\rho+\beta)}\right)\cdot\frac{\vec{\boldsymbol{A}} - \vec{\boldsymbol{O}}_{cyclo}}{\left\|\vec{\boldsymbol{A}} - \vec{\boldsymbol{O}}_{cyclo}\right\|} \qquad \text{Equation 19}$$

The perceived aperture width is therefore the Euclidean distance between the two aperture poles:

$$\hat{W} = \left\|\vec{\hat{\boldsymbol{A}}}_{\boldsymbol{L}} - \vec{\hat{\boldsymbol{A}}}_{\boldsymbol{R}}\right\| \qquad \text{Equation 20}$$

Because the perceptual threshold $W$ reported in Figure 3a was measured in the visual environment, it provides information about the location of the original aperture $\vec{\boldsymbol{A}}$ and its corresponding visual angle $\rho$. This leaves the vergence offset $\beta$ as the only unknown variable for deriving the perceived aperture width.

**4.2 Geometrical Scaling After the Vergence-Accommodation Conflict**

Based on previous studies (Wang et al., 2024b, 2026), the vergence offset associated with the VIVE Pro HMD was approximately 0.22°. This value was used to derive the perceived aperture width based on the perceptual judgment thresholds and Equation 19 and Equation 20. These perceptual thresholds that were adjusted for the VAC were submitted to the same one-factor repeated measures ANOVA (Task 4 levels: UR Pre-VR, VR Pre-Action, VR Post-Action, UR Post-VR). This analysis showed a significant effect of Task on the perceptual thresholds, $F(2.20, 129.89) = 83.93, p < 0.001, \eta_p^2 = 0.59$. Post hoc pairwise comparisons showed significant differences between all task pairs (Figure 6a). Importantly, there was still a difference between the UR Pre-VR and VR Pre-Action (mean difference = 9.48 cm, SE = 0.71, $p < 0.001$), between the VR Pre-Action and Post-Action (mean difference = 2.61 cm, SE = 0.75, $p < 0.01$), and between the UR Post-VR and VR Post-Action (mean difference = 4.36 cm, SE = 0.48, $p < 0.001$). Compared to Figure 3a, it is clear that the VAC adjustment effectively reduced the perceptual thresholds in VR by 5 to 6 cm, but this still did not remove the significant difference between perceptual thresholds in UR and VR.

a

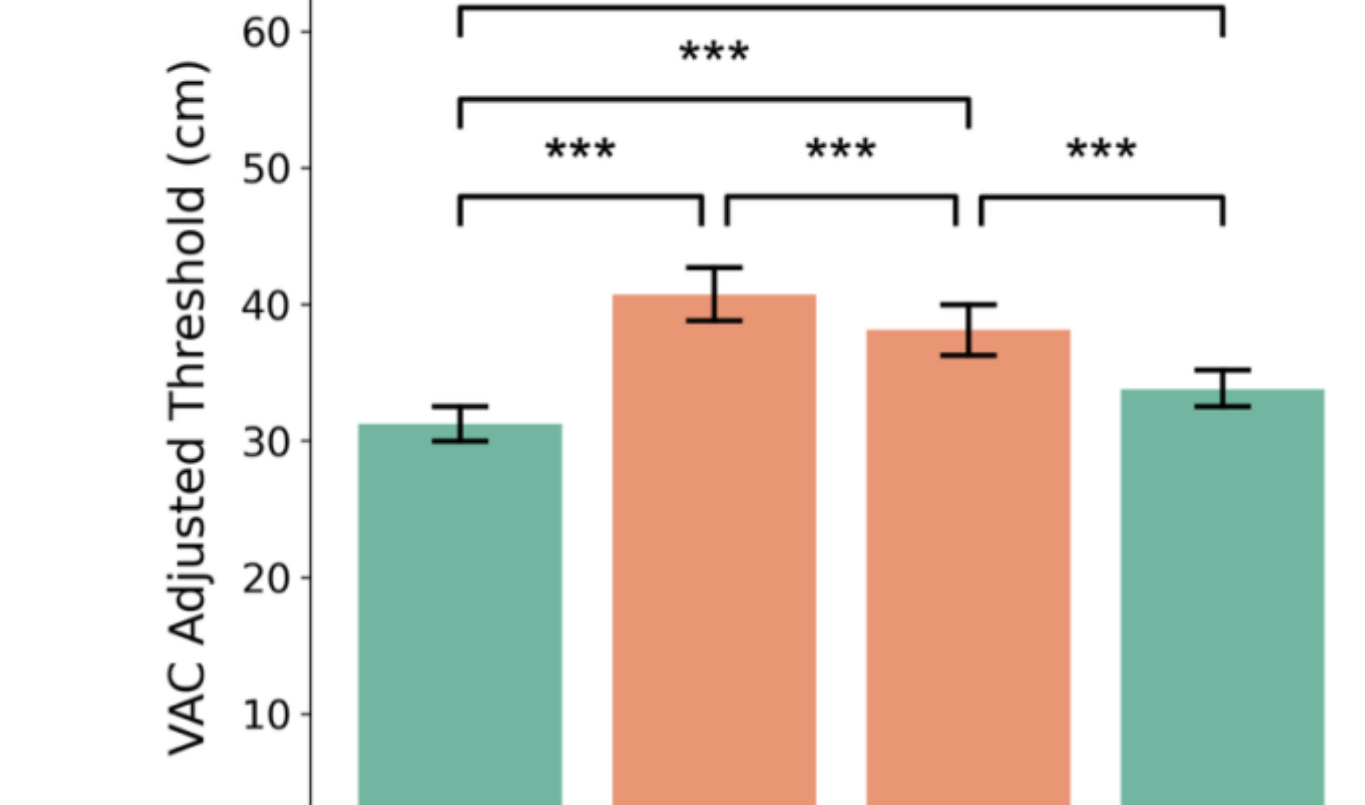

b

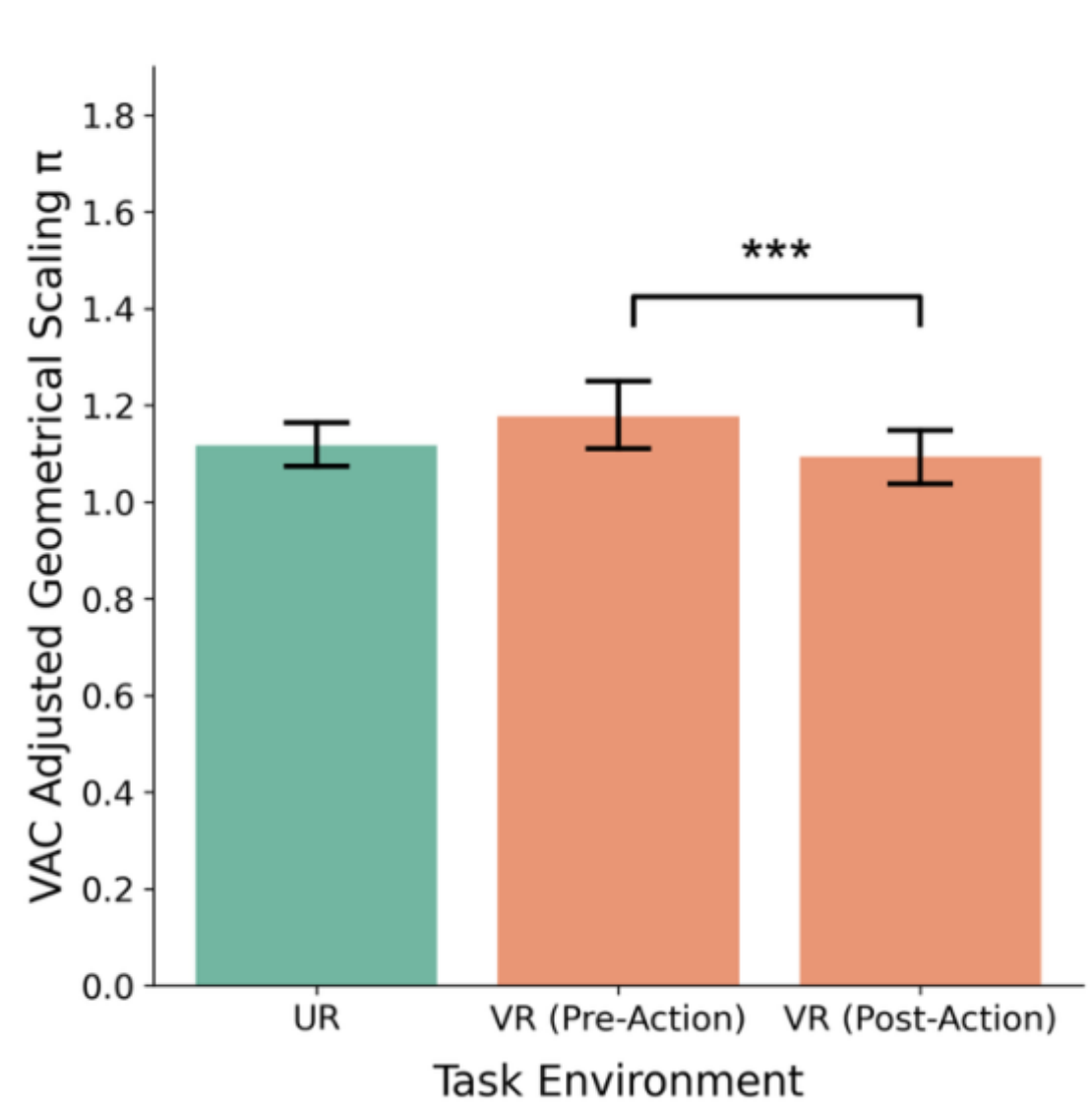

Figure 6. The vergence-accommodation conflict adjusted (a) perceptual threshold and (b) geometrical scaling. Error bars represent the 95% confidence intervals. *: $p < 0.05$; **: $p < 0.01$; ***: $p < 0.001$.

As reasoned earlier, HMD-based VR imposes numerous sensorimotor perturbations onto its users, resulting in heightened uncertainty when performing motor tasks (Wang et al., 2025). In the present experiment, this difference was reflected in elevated action thresholds in the virtual environment compared to UR. If perception remains functionally attuned to action under such uncertainty, perceptual judgments should likewise reflect the altered action boundaries, while preserving the lawful scaling between affordances (perceptual thresholds) and action capabilities (action thresholds). Accordingly, the perception-action scaling relations were quantified for both environments using the VAC adjusted perceptual thresholds (Figure 6b).

A one-factor repeated-measures ANOVA (Task 3 levels) on the geometrical scaling in UR and VR Pre- and Post-Action tasks showed a significant effect of Task, $F(1.65, 97.30) = 5.24, p < 0.05, \eta_p^2 = 0.082$. Post hoc pairwise comparisons only showed a statistically significant difference between the VR Pre-Action and VR Post-Action (mean difference = 0.083, SE = 0.021, $p < 0.001$). This pattern of effects indicates that performing the action in VR facilitated perceptual-motor calibration, enabling participants to attune to the body-scaled affordance of the aperture. As a result, participants reduced the safety margin in their perceptual judgments, reflecting improved alignment between perceived and actual action boundaries within the virtual environment. Importantly, there was no significant difference between the UR and VR Pre-Action (mean difference = 0.059, SE = 0.031, $p = 0.15$) or between the UR and VR Post-Action (mean difference = 0.024, SE = 0.026, $p = 0.64$). Follow-up equivalence testing with paired sample t-tests confirmed statistical equivalence between UR and VR Pre-Action (lower bound: $t(59) = 14.03, p < 0.001$; upper bound: $t(59) = -17.81, p < 0.001$; Hedge's $g = -0.24$, 90% CI = $[-0.45, -0.03]$) and between UR and VR Post-Action (lower bound: $t(59) = 20.33, p < 0.001$; upper bound: $t(59) = -18.50, p < 0.001$; Hedge's $g = 0.12, 90\%$ CI $= [-0.09, 0.33]$).

Overall, these findings suggest that after accounting for VAC and its effect on the perceived width of the aperture, the geometrical scaling between the avatar's body dimensions (i.e., abdominal girth) and the virtual aperture aligned with the scaling relationship observed in UR. In other words, although action in VR was more variable, likely due to sensorimotor perturbations that elevated both action and perceptual thresholds, participants also experienced a separate perceptual distortion that caused the aperture to appear narrower than depicted. This distortion, in turn, affected participants' perception of their functional body size, leading them to judge that a wider aperture was needed for passage. After correcting for this distortion, affordance judgments in VR became consistent with those in the physical world, indicating that participants were able to recalibrate their perception and recover body-scaled affordance invariance in the virtual environment.

## 5. General Discussion

The current study investigated the impact of HMD-based VR on perceptual body size employing a passable aperture (body-based affordance) paradigm. Participants were asked to either sidle through an aperture (action task) or adjust the width of an aperture so that they could sidle through without touching the aperture (perception task). Results showed a significant increase in the action threshold from UR to VR, potentially reflecting the increased uncertainties associated with action performance in VR due to sensorimotor perturbations such as a restricted field of view or a lack of proprioceptive feedback. The perceptual threshold also manifested a similar increase from UR to VR, where experience performing the action task decreased the perceived passable aperture, but not to levels observed in the baseline measures in UR. The increased perceived functional body size in VR transferred to subsequent perceptual judgment in UR. Follow-up analysis derived the affordance ratios, or the geometrical scaling between the perceived critical aperture width (perception threshold) and the participant's action capability for sidling through an aperture (action threshold). Although the affordance ratios should remain invariant for the same task across different environments, the ratios were greater in VR than in UR, even after action calibration. This finding indicated that the increase in the perceptual threshold in VR was disproportionately larger than the increase in the action threshold. Geometrical modelling showed that the additional increase in the perceptual threshold could be attributed to the depth compression due to VAC, which rendered the perceived aperture to be narrower than how the aperture was depicted. After correcting for VAC's effect on perceptual thresholds, the affordance ratios in the UR and VR became comparable, indicating that the invariant geometrical scaling was recovered across modalities. Overall, these data indicate that: 1) action capabilities are impaired in VR due to increased sensorimotor uncertainties; 2) the perceived functional body size increases from UR to VR; 3) this increase transfers from the VR to UR; and 4) this increase can be accounted for by, in large part, VAC in VR. These conclusions will be addressed in turn.

The first key finding from the present study was the increase in action threshold from UR to VR, indicating reduced action capabilities of sidling through an aperture. HMD-based VR imposes significant sensorimotor perturbations on its users, such as a restricted field of view (Gagnon et al., 2021), degraded or the absence of proprioceptive (Mestre et al., 2016) and haptic feedback (Brock et al., 2023), and motion-to-photon latency (Warburton et al., 2023). These perturbations may reduce action accuracy and increase action variability (Wang et al., 2025). For instance, Brock et al. (2023) compared the kinematics of golf putting in UR and VR and showed increased magnitude and variability of postural movement in VR. In the context of the current study, such increased movement variability potentially made it harder for participants to modulate their torso position relative to the aperture, thereby shifting the action boundary outward for larger safety margins. Theories in motor learning posit that successful learning entails reducing task-irrelevant variability while preserving or exploiting task-relevant variability (Dhawale et al., 2017; Todorov & Jordan, 2002). For VR-based skill training intended to transfer to UR, if VR increases task-relevant movement variability due to task-irrelevant sensorimotor perturbations, participants must learn to control these effects during action performance. This compensation introduces additional, device-specific learning demands that may not generalize to performance in UR, potentially reducing the effectiveness of using VR as a motor training tool.

The second main finding is the consistent increase in perceptual thresholds from UR to VR. This finding indicates that participants perceived their virtual body as larger relative to the surrounding environment on a functional level. Geometrical modelling revealed that this increase was in part due to the reduced action capabilities in VR and in part due to the depth compression resulting from VAC. Previous studies have performed a similar comparison of performance between UR and VR in a passable aperture task (Bhargava, Lucaites, et al., 2020; Bhargava, Solini, et al., 2020). Bhargava, Lucaites, et al. (2020) compared passability judgment of walking through a doorway in a physical and an immersive virtual environment, but did not find a difference between the modalities. However, the setup of the Bhargava et al. study allowed participants to move closer to the aperture if they were uncertain of their judgment. As a result, the viewing distance in VR (mean = 134 cm) was noticeably shorter than that in UR (mean = 190 cm); note that the viewing distance in the present study was kept consistent at 2.5 m in both UR and VR. Bhargav et al. argued that the shorter viewing distance in VR provided participants with additional dynamic visual information as they walked to the desired location, thereby reducing the uncertainty in the passability judgment and resulting in perceptual thresholds that were not different from those in UR.

The viewing distance is an important factor to consider because this distance mediates the effect of depth compression due to the vergence offset (Figure 5), which would in turn affect the perceived aperture width. Figure 7 illustrates the effect of viewing distance on perceived distance and width of an aperture based on a depicted width of 30 cm. As the viewing distance decreases, the amount of depth compression diminishes, and the perceived aperture width becomes closer to the depicted width. As a result, by allowing participants to be closer to the aperture in VR, Bhargava, Lucaites, et al. (2020) effectively reduced the magnitude of perceptual distortions on the perceived aperture, reducing the difference in perceptual thresholds between the physical and virtual environments. In fact, in a follow-up study, Bhargava, Solini, et al. (2020) fixed the viewing distance of 2 m and found a much higher perceptual threshold in VR compared to in UR. This result is consistent with the findings from the current study. This result also reinforces the role of depth compression, which is mediated by viewing distance, as a key driver of increased perceptual thresholds in VR. When the viewing distance is fixed and extended, as in both the present study and Bhargava, Solini et al.'s (2020) follow-up study, participants are more susceptible to the effects of VAC-induced distortion, which causes apertures to appear narrower than depicted. This narrower perception of the apertures, in turn, leads to an overestimation of functional body size relative to the environment, resulting in increased perceptual thresholds. These converging findings underscore the importance of viewing geometry in modulating how VR alters body-scaled affordances and perceived spatial layout.

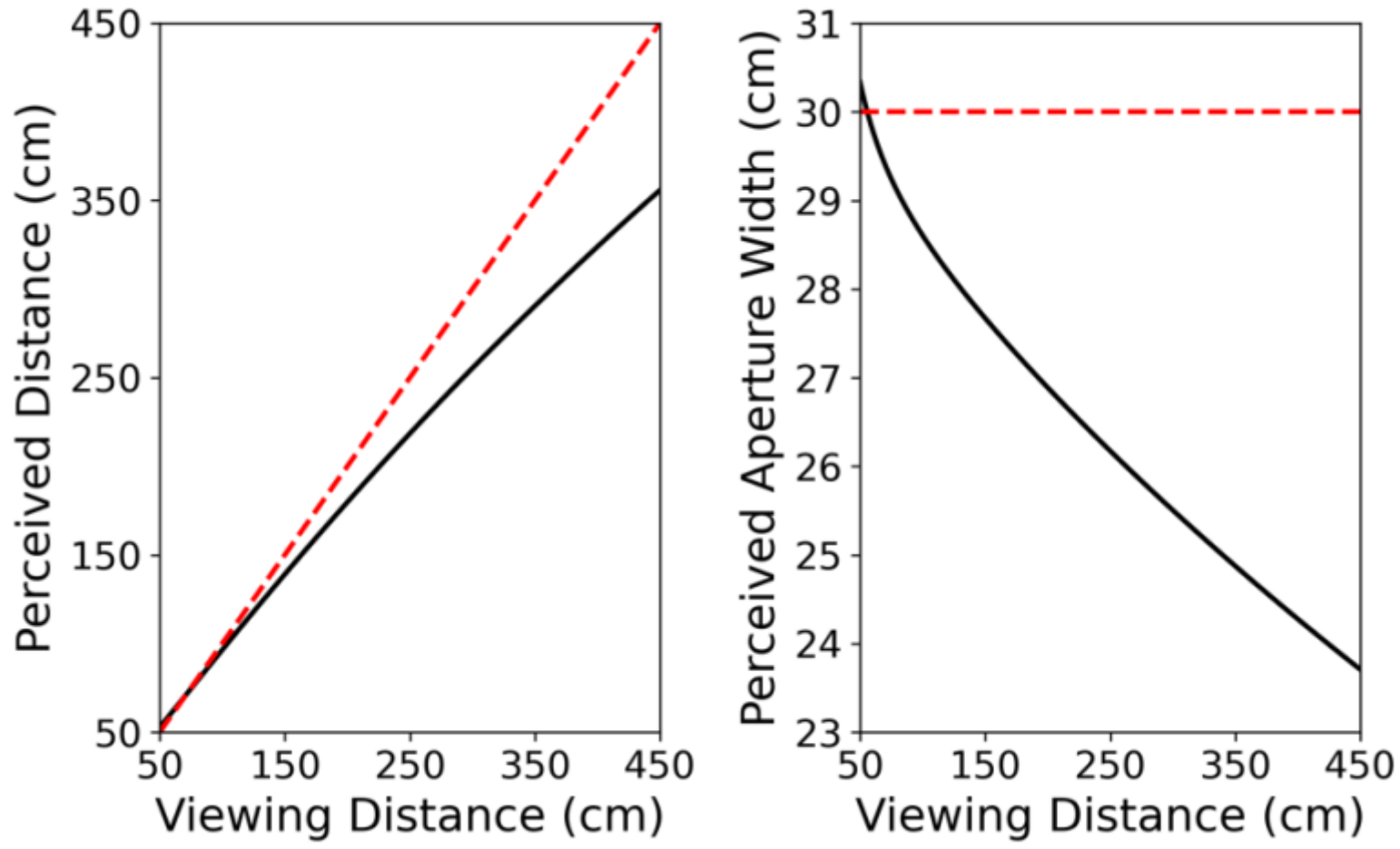

Figure 7. An illustration of the effect of the vergence offset (0.22°) and viewing distance on perceived aperture distance (left) and perceived aperture width (right) for an aperture with a width of 30 cm. The red dashed lines indicate veridical distance and aperture width.

The third major finding was that the underestimation of aperture width and the resulting overestimation of functional body size experienced in VR persisted when they returned to the physical world after experience in the VR environment. A similar adaptation aftereffect when returning to UR following VR exposure was observed in a study of manual pointing tasks (Wang et al., 2024a). In Wang et al. (2024a), participants pointed at the same series of targets in the physical environment before and after pointing to these targets in a virtual environment. Compared to their pre-VR performance in UR, participants noticeably undershot targets in VR and again in their post-VR performance in UR. This undershooting aftereffect in UR gradually diminished over time. The authors attributed the undershooting in VR to the depth compression due to VAC. They further suggested that the aftereffect emerged because the visuomotor system adapted to the fixed accommodation distance imposed by VR HMD, which disrupted the accommodative vergence process and resulted in a lingering vergence offset after leaving the immersive VR. In the current study, there was an approximate 8% (2.52 cm / 31.23 cm) increase in perceptual threshold in UR from pre- to post-VR performance, congruent with the overestimation in VR. This finding suggests that not only did people perceive the aperture to be smaller relative to their body in VR, but this effect also transfers back to their perception of their functional body size and movement capabilities in UR after exposure to VR.

In addition to the VAC-related adaptation aftereffect, another type of carryover effect relevant to the current study concerns affordance calibration and recalibration (Bingham, Pan, et al., 2014; Bingham & Pagano, 1998; Fajen, 2008; Franchak, 2020; Pan et al., 2014; Rieser et al., 1995; Withagen & Michaels, 2002). From an ecological perspective, affordances specify action possibilities in organism-scaled units (J. J. Gibson, 1979), but effectively using this information requires calibration, or learning the mapping between perceptual information and one's action capabilities through feedback and practice (Bingham & Pagano, 1998). In the context of sidling through an aperture, observers need to relate visual information specifying aperture width to a relevant body dimension (i.e., abdomen girth) to judge and control if successful passage is possible (e.g., Franchak & Adolph, 2014), and this mapping is refined through action experience and outcome feedback (Franchak, 2017). In the current study, participants first gained action experience in UR and then performed the same task in VR, where HMD-induced sensorimotor perturbations increased the action threshold (i.e., wider apertures were required), consistent with a recalibration of effective action capabilities in VR. A natural extension of this calibration framework is that such recalibration could carry over to subsequent action performance in UR. However, because the present design did not include a post-VR UR action block, it cannot be determined whether VR exposure altered the UR action threshold (and an affordance ratio computed entirely within UR). Future work should include post-VR UR action measurements to test whether the perceptual aftereffects observed here are accompanied by corresponding carryover changes in action capabilities and safety margins in UR.

A final important finding of the present work was the effectiveness of the geometrical VAC model in predicting the perceived aperture width. It would also be interesting to examine the magnitude of depth compression in VR using this model. Equation 19 can not only be used to compute aperture width, but also provides information about the distance of the aperture from the observer along the depth dimension. Based on the perceptual thresholds in VR, the average perceived distance between the aperture and the observer was approximately 219 cm for both the Pre- and Post-Action perception tasks. This value corresponds to a depth compression ratio of 0.88 at a viewing distance of 2.5 m. A previous study that investigated the perceptual geometry of VR using an exocentric pointing task also reported a similar depth compression ratio of 0.84 (95% confidence interval: ± 0.14) (Wang & Troje, 2023). The relatively consistent depth compression ratios across experiments support the effectiveness of the geometrical VAC model in explaining VR-induced depth compression, which arises from the angular offset added to the vergence angle that disrupts binocular geometry. In fact, previous research has shown that distance estimation based on binocular disparity is effective for distances up to 9 m (Allison et al., 2009). Therefore, extending prior work that focused primarily on visually guided pointing movements within personal space (0 – 2 m) (Wang et al., 2024b, 2026), the current study demonstrated that VAC associated with VR HMDs also affects perception and action in the broader action space (Cutting & Vishton, 1995). This finding challenges the prevailing view that VAC only affects visual perception in the personal space (Renner et al., 2013). Although convergence and accommodation may not specify depth effectively beyond personal space, VAC perturbs binocular disparity that translates to distortions in perceived depth beyond personal space.

In sum, this study demonstrates that there are perceptual distortions induced in VR by the HMD. These distortions extend beyond personal space into the action space, altering perception when returning to the physical world. By integrating behavioral affordance measures with geometrical modelling, this study reveals how VAC systematically compresses perceived depth and alters perceived body-environment relationships. These effects not only manifest during VR interaction but also persist afterward, suggesting a recalibration of visual-motor coupling. Future research should explore whether repeated VR exposure strengthens or attenuates these aftereffects and investigate potential strategies, such as dynamic calibration of avatar size or adaptive optics with shader programs (Wang et al., 2026), to mitigate depth distortions in VR. Additionally, VR shifted the action boundary outward relative to UR, indicating larger safety margins and execution uncertainty under HMD-induced sensorimotor perturbations. These device-specific uncertainties impose learning demands that may not generalize, hindering transfer. Future work should quantify the contribution of specific perturbations and develop mitigation strategies, either by reducing them (e.g., wider field of view, improved haptics/latency) or by training compensation, to improve VR to UR motor skill transfer.

**Funding**

This project draws upon research supported by the Government of Canada’s New Frontiers in Research Fund (NFRF). We also acknowledge the support of the Natural Sciences and Engineering Research Council of Canada (NSERC).

**Conflict of Interests**

The authors have no conflicts of interest to declare that are relevant to the content of this article.

**Data Availability**

The datasets generated during and/or analyzed during the current study are not publicly available due to privacy concerns but are available from the corresponding author upon reasonable request.

**Acknowledgement**

The authors would like to thank Aarohi Pathak, Lauren Hong, Zoe Lin, and Eva Huang for assisting with the data collection. The author would also like to thank Geoffrey Bingham for his thoughtful comments and feedback.